\documentclass[12pt,a4paper]{article}
\usepackage{epsfig}
\begin{document}
\title{ New gauge boson $B_{H}$ production associated with W boson pair via $\gamma\gamma$ collision
in the littlest Higgs model}

\author{Xuelei Wang$^{a,b}$, Jihong Chen$^{b}$,Yaobei Liu$^{b}$,Suzhen Liu$^{b}$,Hua Yang$^{c}$   \\
 {\small a: CCAST (World Laboratory) P.O. BOX 8730. B.J. 100080 P.R. China}\\
 {\small b: College of Physics and Information Engineering,}\\
 \small{Henan Normal
University, Xinxiang  453007. P.R.China}
\thanks{This work is supported by the National Natural Science
Foundation of China(Grant No.10375017 and No.10575029).}
\thanks{E-mail:wangxuelei@sina.com}\\
{\small c: Department of Mathematics and Physics, College of
Science, }\\
{\small Information Engineering University, Zhengzhou
450001,P.R.China.}}
 \maketitle
\begin{abstract}
\indent The littlest Higgs model is the most economical little
Higgs model. The observation of the new gauge bosons predicted by
the littlest Higgs model could serve as a robust signature of the
model. The ILC, with the high energy and luminosity, can open an
ideal window to probe these new gauge bosons, specially, the
lightest $B_H$. In the framework of the littlest Higgs model, we
study a gauge boson $B_{H}$ production process
$\gamma\gamma\rightarrow W^{+}W^{-}B_H$. The study shows that the
cross section of the process can vary in a wide
range($10^{-1}-10^1$ fb) in most parameter spaces preferred by the
electroweak precision data. The high c.m. energy(For example,
$\sqrt{s}=1500$ GeV) can obviously enhance the cross section to
the level of tens fb. For the favorable parameter spaces, the
sufficient typical events could be assumed at the ILC. Therefore,
our study about the process $\gamma\gamma\rightarrow
W^{+}W^{-}B_H$ could provide a useful theoretical instruction for
probing $B_H$ experimentally at ILC. Furthermore, such process
would offer a good chance to study the triple and quartic gauge
couplings involving $B_H$ and the SM gauge bosons which shed
important light on the symmetry breaking features of the littlest
Higgs model.
\end{abstract}
PACS number(s): 12.60Nz,14.80.Mz,12.15.Lk,14.65.Ha

\newpage
\section{Introduction}
\hspace{0.6cm} At present the success of the standard model(SM) is
well known and doubtless. However, the mechanism of electroweak
symmetry breaking(EWSB) remains the most prominent mystery in
current particle physics. The Higgs particle that is assumed to
trigger the EWSB in the SM has not been found. In addition, there
are prominent problems of triviality and unnaturalness in the
Higgs sector. Thus, the SM can only be an excellent effective
field theory below some high energy scales. New physics beyond the
SM should exist at the TeV scale. The possible new physics
scenarios at the TeV scale might be
supersymmetry(SUSY)\cite{supersymmetry}, dynamical symmetry
breaking\cite{dynamical}, extra dimensions\cite{dimensions}, the
little
 Higgs model \cite{little-1,little-2,little-3,littlest} etc.

 \indent  Among the extended models beyond the SM, the little
Higgs model offers a very promising solution to the hierarchy
problem in which the Higgs boson is naturally light as a result of
nonlinearly realized symmetry. The key feature of this model is
that the Higgs boson is a pseudo-Goldstone boson of an approximate
global symmetry which is spontaneously broken by a vacuum
expectation value(vev) at a scale of a few TeV and thus is
naturally light. Such model can be regarded as the important
candidate of new physics beyond the SM. The littlest Higgs  model
\cite{littlest} is a simplest and phenomenologically viable model
to realizes the little Higgs idea. The model predicts the presence
of the new gauge bosons $(B_{H},Z_{H},W^{\pm}_{H})$ and their
masses are in the range of a few TeV, except for $B_{H}$ in the
range of hundreds of GeV. The minimality of the littlest Higgs
model would leave the characteristic signatures at present and
future high energy collider experiments.

\indent It is widely believed that the hadron colliders, such as
the Tevatron and the LHC running in 2007, can directly probe the
possible new physics beyond the SM up to a few TeV, while the TeV
energy linear $e^{+}e^{-}$ collider(LC) is also required to
complement the probe of the new particles with detailed
measurements\cite{LC}. A unique feature of the TeV energy LC is
that it can be transformed to $\gamma\gamma$ or $e\gamma$
collider(the photon collider) by the laser-scattering method. In
this case the energy and luminosity of the photon beams would be
the same order of magnitude of the original electron beams and the
set of final states at the photon collider is much rich than that
in the $e^{+}e^{-}$ mode. Furthermore, one can vary polarizations
of photon beams relatively easily, which is advantageous for
experiments. In some scenarios, the photon collider is the best
instrument for the discovery of the signals of the new physics
models and will be able to study multiple vector boson production
with high precision.

\indent The probe of the new particles, specially the new gauge
bosons predicted by the littlest Higgs model, can provide a direct
way to test the model. The CERN Large Hadronic Collider(LHC), with
the center of energy $\sqrt{s}=14$ TeV, has the ability to produce
these heavy new particles. In some literatures\cite{Han,LHC}, the
production mechanism of these new particles at the LHC has been
studied and the most promising production process is Drell-Yan
process which shows that the LHC has the potential to detect them.
However, the detailed study of these new gauge boson couplings
needs the precision measurement at the future LC, and such work
can be performed at the planned International Linear Collider(ILC)
with the center of mass(c.m.) energy $\sqrt{s}$=300 GeV-1.5 TeV
and the yearly luminosity 500 $fb^{-1}$\cite{ILC}. Specially for
the gauge boson $B_{H}$, we find that the global symmetry
structure $SU(5)/SO(5)$ of the littlest Higgs model allows a
substantially light $B_{H}$ with the mass about a few hundred GeV,
and such gauge boson is light enough to be produced at the first
running of the ILC. Therefore, the exploring of this light $B_{H}$
at the ILC would play an important role in testing the littlest
Higgs model. Some phenomenological studies of littlest Higgs model
via $e^+e^-$ collision at the ILC has been done\cite{ILCstudy}.
 The ILC has the capability to discover the effects of the littlest
 Higgs model over the entire
 theoretically interesting range of parameters and to determine the
 couplings of the heavy gauge bosons to
 the precision of a few percent. The realization of $\gamma\gamma$
 collision would provide us a better
chances to probe $B_{H}$. In this paper, we study a $B_{H}$
production process associated with W boson pair via $\gamma\gamma$
collision realized at the ILC, i.e., $\gamma\gamma\rightarrow
W^+W^-B_H$. The motivation to study this process is that
$W^{+}W^{-}B_{H}$ production mode can be realized below the TeV
scale and its cross section is large enough to detect $B_{H}$ with
the high energy and luminosity of the ILC. On the other hand, such
process offers a direct study of vector boson couplings in the
littlest Higgs model.

 \indent  This paper is organized as follows. In
the section two, we briefly review the littlest Higgs model. In
the section three, the calculation process of the cross section is
presented. The section four contains our numerical results and
conclusions.

\section{Brief review of the  littlest Higgs model}

\hspace{0.6cm} The minimal model containing the little Higgs ideas
is called the littlest Higgs model. This model is based on a
non-linear sigma model and consists of a global $SU(5)$ symmetry
which is broken down to $SO(5)$ by a vacuum condensate
$f\sim\frac{\Lambda_{s}}{4\pi}\sim$ TeV which results in 14
Goldstone bosons. The effective field theory of those Goldstone
bosons is parameterized by a non-linear $\sigma$-model with a
gauge symmetry $[SU(2)\times U(1)]^{2}$. The breaking of the
global $SU(5)$ down to $SO(5)$ simultaneously breaks $[SU(2)\times
U(1)]^{2}$ down to its diagonal $SU(2)_L\times U(1)_Y$ subgroup,
which is identified as the SM electroweak gauge group. In
particular, the Lagrangian will still preserve the full
$[SU(2)\times U(1)]^{2}$ gauge symmetry.

The leading order dimension-two term in the non-linear
$\sigma$-model can be written for the scalar sector as
\begin{eqnarray}
 {\pounds}_{\Sigma}=\frac{f^2}{8}Tr|D_{\mu}\Sigma|^2,
\end{eqnarray}
with the covariant derivative
\begin{eqnarray}
 D_{\mu}\Sigma=\partial_{\mu}\Sigma-i\sum^{2}_{j=1}[g_j(W_j\Sigma+\Sigma W^T_j)+
 g'_j(B_j\Sigma+\Sigma B^T_j)].
\end{eqnarray}
Where $W_j=\sum^3_{a=1}W^a_{\mu j}Q^a_j$ and $B_j=B_{\mu j}Y_j$
are the $SU(2)$ and $U(1)$ gauge fields, respectively. The
low-energy dynamics is described in term of the non-linear sigma
model field
\begin{eqnarray}
 \Sigma(x)=e^{2i\Pi/f}\Sigma_0,
\end{eqnarray}
where $\Pi=\sum _a\pi^a(x)X^a$. The sum runs over the 14 broken
$SU(5)$ generators $X^a$, and $\pi^a(x)$ are the Goldstone bosons.
The symmetry breaking vev is proportional to $\Sigma_0$.

  The gauge boson mass eigenstates after
the  spontaneous gauge symmetry breaking are
\begin{eqnarray}
W=sW_1+cW_2, ~~~~W'=-cW_1+sW_2,\\
B=s'B_1+c'B_2,~~~~B'=-c'B_1+s'B_2.
\end{eqnarray}
$s,c,s',c'$ are mixing angles and are defined as
\begin{eqnarray}
s\equiv sin\psi=\frac{g_2}{\sqrt{g^2_1+g^2_2}},~~~~s'\equiv
sin\psi^{\prime}=\frac{g'_2}{\sqrt{g'^2_1+g'^2_2}}, \\
c\equiv cos\psi=\sqrt{1-s^2},~~~~c'\equiv
cos\psi^{\prime}=\sqrt{1-s'^2}.
\end{eqnarray}
The $W$ and $B$ remain massless and are identified as the SM gauge
bosons, with couplings
\begin{eqnarray}
g=g_1s=g_2c, \ \ \  g'=g'_1s'=g'_2c'.
\end{eqnarray}
$W'$ and $B'$ are the heavy the massive gauge bosons associated
with the four broken generators of $[SU(2)\times U(1)]^{2}$.

After linearize the theory, the couplings of the gauge bosons to
Higgs field can be obtained via Eq.(1) at leading order in $1/f$
\begin{eqnarray}
\frac{1}{4}H[g^2(W^{\mu a}W^a_{\mu}-W'^{\mu a}W'^a_{\mu}-2cot2\psi
W'^{\mu a}W^a_{\mu})+g'^2(B^{\mu}B_{\mu}-B'^{\mu
}B'_{\mu}-2cot2\psi^{\prime} B'^{\mu}B_{\mu})]H^{\dag}.
\end{eqnarray}
where $H=(h^+,h_0)$ is the SM Higgs doublet.

In the SM, the four-point couplings of the form $W^{a}_{\mu}W^{\mu
a}H^\dag H$ and $B_{\mu}B^{\mu}H^\dag H$ lead to a quadratically
divergent in the Higgs mass arising from the seagull diagrams
involving gauge boson loops. From Eq(9), we can see that, in the
littlest Higgs model, the $W'^{a}_{\mu}W'^{\mu a}H^\dag H$ and
$B'_{\mu}B'^{\mu}H^\dag H$ couplings have unusual forms which
serves to exactly cancel the quadratic divergence in the Higgs
mass leaded by $W^{a}_{\mu}W^{\mu a}H^\dag H$ and
$B_{\mu}B^{\mu}H^\dag H$. The cancellation of such divergence is a
crucial feature of the little Higgs theory. The key test of the
little Higgs mechanism in the gauge sector is the experimental
verification of this feature. Some literatures have discussed the
prospects at the LHC\cite{LHC}.

 The EWSB in the littlest Higgs model is triggered by the
Higgs potential generated by one-loop radiative correction and
massless $W$ and $B$ obtain their masses. The Higgs potential
includes the parts generated by the gauge boson loops as well as
the the fermion loops.

 The EWSB induces further mixing between the light and heavy gauge bosons and  the final
 observed mass eigenstates are the light SM-like bosons $W_L^{\pm},Z_L$ and
$A_L$ observed in experiment, and new heavy bosons $W^{\pm}_H,Z_H$
and $B_H$ that could be observed in future experiments. The masses
of neutral gauge bosons are given to $O(v^2/f^2)$
by\cite{ILCstudy}
\begin{eqnarray}
M^{2}_{A_{L}}&=&0,\\
M^{2}_{Z_{L}}&=&(M^{SM}_{Z})^{2}\{1-\frac{v^{2}}{f^{2}}[\frac{1}{6}+\frac{1}{4}(c^{2}-s^{2})^{2}+
\frac{5}{4}(c'^{2}-s'^{2})^{2}]+8\frac{v'^2}{v^2}\},\\
M^{2}_{Z_{H}}&=&(M^{SM}_{W})^{2}\{\frac{f^2}{s^2c^2v^2}-1+\frac{v^2}{2f^2}[\frac{(c^2-s^2)^2}{2c_W^2}
+\chi_H\frac{g'}{g}\frac{c'^2s^2+c^2s'^2}{cc'ss'}]\},\\
M^{2}_{B_{H}}&=&(M^{SM}_{Z})^{2}s^{2}_{W}\{\frac{f^{2}}{5s'^2c'^2v^2}-1
+\frac{v^2}{2f^2}[\frac{5(c'^2-s'^2)^2}{2s^2_W}-\chi_H\frac{g}{g'}\frac{c'^2s^2+c^2s'^2}{cc'ss'}]\}.
\end{eqnarray}
Where
$\chi_{H}=\frac{5}{2}gg'\frac{scs'c'(c^{2}s'^{2}+s^{2}c'^{2})}{5g^{2}s'^{2}c'^{2}-g's^{2}c^{2}}$,
$v$=246 GeV is the elecroweak scale, $v'$ is the vev of the scalar
$SU(2)_{L}$ triplet and $s_{W}(c_{W})$ represents the sine(cosine)
of the weak mixing angle.

On the other hand, the EWSB also induces cubic couplings between
the physical Higgs boson and the gauge bosons. The explicit forms
of these couplings can be obtained from Eq.(1) after linearizing
the theory. The three diagonal coupling, $hZZ,~hZ_HZ_H,~hB_HB_H$,
add up to zero. So do the two diagonal couplings $hW^+W^-$ and
$hW^+_HW^-_H$. These cancellations can be traced back to Eq.(9),
and are therefore directly related to the crucial feature of
cancellation of quadratic divergences. Measuring the diagonal
coupling would provide the most direct way to verify the little
Higgs theory.  Such measurement requires associated production of
a new heavy boson with a Higgs and is a difficult task. However,
it is much easier to measure the off-diagonal couplings, such as
$hZ_HZ,~hB_HZ,~hW^{\pm}_HW^{\mp}$. Although these couplings do not
directly participate in the cancellation of quadratic divergences,
verifying their structure would provide a strong evidence for the
crucial feature of the model.

The gauge kinetic terms $\pounds_{G}$ take the standard form:
\begin{eqnarray}
\pounds_{G}=-\frac{1}{4}\sum^{2}_{j=1}(W^{\mu\nu}_{ja}W^a_{j\mu\nu}+B^{\mu\nu}_{ja}B^a_{j\mu\nu}).
\end{eqnarray}
These terms yield 3- and 4-particle interactions among the gauge
bosons.

There are still other interactions in the littlest Higgs model,
${\pounds}_{F}, {\pounds}_{Y},V_{CW}$.  The fermion kinetic terms
${\pounds}_{F}$ can give the couplings of the gauge bosons with
fermions. The couplings of the scalars h and $\phi$ with fermions
can be derived from the Yukawa interaction terms ${\pounds}_{Y}$.
 The effective Higgs potential, the Coleman-weinberg potential
$V_{CW}$, is generated at one-loop and higher orders due to the
interactions of Higgs with gauge bosons and fermions, which can
induce the EWSB by driving the Higgs mass squared parameter
negative.

It is shown via above discussion that the new heavy gauge bosons
$B_H,Z_H,W^{\pm}_H$ are predicted in the littlest Higgs model and
these new particles might produce the characteristic signatures at
the present and future high energy collider experiments
\cite{Han,LHC,ILCstudy,signiture}.

\section{The cross section of the process $\gamma\gamma\rightarrow W^{+}W^{-}B_H$}
 \indent From the gauge kinetic terms $\pounds_{G}$, one can derive the
3-point and 4-point gauge self-coupling expressions. With all
momenta out-going, the 3-point gauge boson self-couplings can be
written in form of\cite{Han}
\begin{eqnarray}
V_1^{\mu}(k_1)V_2^{\nu}(k_2)V_3^{\rho}(k_3):~~
-ig_{V_1V_2V_3}[g^{\mu\nu}(k_1-k_2)^{\rho}+g^{\nu\rho}(k_2-k_3)^{\mu}+g^{\rho\mu}(k_3-k_1)^{\nu}],
 \end{eqnarray}
and the 4-point gauge boson self-couplings take the form
\begin{eqnarray}
W_1^{+\mu}W_2^{+\nu}W_3^{-\rho}W_4^{-\sigma}:~~
-ig_{W_1^+W_2^+W_3^-W_4^-}(2g^{\mu\nu}g^{\rho\sigma}-g^{\mu\rho}g^{\nu\sigma}-
g^{\nu\rho}g^{\mu\sigma}),\\
V_1^{\mu}V_2^{\nu}W_1^{+\rho}W_2^{-\sigma}:~~
ig_{V_1V_2W_1^+W_2^-}(2g^{\mu\nu}g^{\rho\sigma}-g^{\mu\rho}g^{\nu\sigma}-
g^{\nu\rho}g^{\mu\sigma}).
 \end{eqnarray}
The coefficients $g_{V_1V_2V_3},~ g_{W_1^+W_2^+W_3^-W_4^-}$ and
$g_{V_1V_2W_1^+W_2^-}$ are given as

\begin{eqnarray}
g_{A_LW^+_LW^-_L}=-e,\ \ \
 g_{W^+_LW^-_LB_H}=\frac{ec_W}{s_W}\frac{v^2}{f^2}x_{Z}^{B'},\\
 g_{A_LA_L W^+_LW^-_L}=-e^2,\ \ \ \ g_{A_L
W^+_LW^-_LB_H}=\frac{e^2c_W}{s_W}\frac{v^2}{f^2}x_{Z}^{B'}.
 \end{eqnarray}
Where $x_{Z}^{B'}=-\frac{5}{2s_W}s'c'(c'^{2}-s'^{2})$. We can see
that the couplings $A_LW^+_LW^-_L$ and $A_LA_L W^+_LW^-_L$ are
just the same as the couplings $\gamma W^+W^-$ and $\gamma\gamma
W^+W^-$ in the SM. Comparing the mass of $W_L$ with the SM $W$, we
find there exists a correction term in order of $v^2/f^2$. Limited
by the electroweak precision data, such correction should be much
small. So, we can safely regard $A_L$ and $W_L$ in the littlest
Higgs model as $\gamma$ and $W$ in the SM. So, we will represent
$A_L,W_L$ as $\gamma,W$ in the following discussion.

Via above 3-point and 4-point gauge self-couplings, the process
$\gamma\gamma\rightarrow W^{+}W^{-}B_H$ can be realized in the way
shown in Fig.1.

The crossing diagrams with the interchange of the two incoming
photons are not shown. The initial photons are denoted by
$\varepsilon_{\mu}(k_{1})$,  $\varepsilon_{\nu}(k_{2})$ and the
final state $W^{+},W^{-},B_{H}$ are given by
$\varepsilon_{\alpha}(p_+),~\varepsilon_{\beta}(p_-),~\varepsilon_{\gamma}(k_3)$,
respectively.  The production amplitudes of the process can be
written as
 \begin{eqnarray}
M&=&G\cdot\varepsilon_{\mu}(k_{1})\varepsilon_{\nu}(k_{2})\varepsilon_{\alpha}(p_{+})\varepsilon_{\beta}(p_{-})
\varepsilon_{\gamma}(k_{3})M^{\mu\nu\alpha\beta\gamma},
 \end{eqnarray}
with
\begin{eqnarray}
M^{\mu\nu\alpha\beta\gamma}&=&\sum_{\zeta=a,b,c,d,e,f,g}M_{\zeta}^{\mu\nu\alpha\beta\gamma},
 \end{eqnarray}
with
\begin{eqnarray}
G=\frac{5}{2}\frac{e^{3}c_{W}}{s_{W}^{2}}\frac{v^{2}}{f^{2}}s'c'(c'^{2}-s'^{2}).
 \end{eqnarray}
 Where $M_{\zeta}^{\mu\nu\alpha\beta\gamma}$ is the contribution of Fig.1(a-g) to the
 process.

\indent In order to write a compact expression for the amplitudes,
it is convenient to define the triple-boson couplings coefficient
as
\begin{eqnarray}
\Gamma_{3}^{\alpha\beta\gamma}(p_{1},p_{2},p_{3})&=&g^{\alpha\beta}(p_{1}-p_{2})^{\gamma}+g^{\beta\gamma}(p_{2}-p_{3})^{\alpha}+g^{\gamma\alpha}(p_{3}-p_{1})^{\beta},
 \end{eqnarray}
 with all momenta out-going, the quartic-boson coupling
 coefficient as
\begin{eqnarray}
\Gamma_{4}^{\mu\nu\alpha\beta}&=&2g^{\mu\nu}g^{\alpha\beta}-g^{\mu\alpha}g^{\nu\beta}-g^{\nu\alpha}g^{\mu\beta},
 \end{eqnarray}
  and the W boson propagator tensor
\begin{eqnarray}
D^{\mu\nu}(k)&=&\frac{g^{\mu\nu}}{k^{2}-M_W^{2}}.
 \end{eqnarray}
 Using the above definitions, we can explicitly write $M_{\zeta}^{\mu\nu\alpha\beta\gamma}$ as
\begin{eqnarray}
&&M_{a}^{\mu\nu\alpha\beta\gamma}=\Gamma_{3}^{\alpha\gamma\xi}(p_{+},k_{3},-p_{+}-k_{3})
D_{\xi\sigma}(p_{+}+k_{3})\Gamma_{3}^{\mu\sigma\rho}(-k_{1},p_{+}+k_{3},k_{2}-p_{-})\\
\nonumber &&\hspace{1.7cm}D_{\rho\lambda}(p_{-}-k_{2})
\Gamma_{3}^{\beta\nu\lambda}(p_{-},-k_{2},p_{-}-k_{2})+[k_{1\leftrightarrow2};\mu\leftrightarrow\nu],\\
\nonumber
&&M_{b}^{\mu\nu\alpha\beta\gamma}=\Gamma_{3}^{\beta\xi\gamma}(p_{-},-p_{-}-k_{3},k_{3})
D_{\xi\sigma}(p_{-}+k_{3})\Gamma_{3}^{\sigma\nu\rho}(p_{-}+k_{3},-k_{2},p_{+}-k_{1})\\
\nonumber &&\hspace{1.7cm}D_{\rho\lambda}(p_{+}-k_{1})
\Gamma_{3}^{\mu\alpha\lambda}(-k_{1},p_{+},k_{1}-p_{+})+[k_{1\leftrightarrow2};\mu\leftrightarrow\nu],\\
\nonumber
&&M_{c}^{\mu\nu\alpha\beta\gamma}=\Gamma_{3}^{\mu\alpha\xi}(-k_{1},p_{+},k_{1}-p_{+})
D_{\xi\sigma}(k_{1}-p_{+})\Gamma_{3}^{\sigma\gamma\rho}(p_{+}-k_{1},k_{3},p_{-}-k_{2})\\
\nonumber &&\hspace{1.7cm} D_{\rho\lambda}(p_{-}-k_{2})
\Gamma_{3}^{\beta\nu\lambda}(p_{-},-k_{2},k_{2}-p_{-})+[k_{1\leftrightarrow2};\mu\leftrightarrow\nu],\\
\nonumber
&&M_{d}^{\mu\nu\alpha\beta\gamma}=\Gamma_{3}^{\beta\nu\xi}(p_{-},-k_{2},k_{2}-p_{-})
D_{\xi\lambda}(k_{2}-p_{-})\Gamma_{4}^{\lambda\alpha\mu\gamma}+[k_{1\leftrightarrow2};\mu\leftrightarrow\nu],\\
\nonumber
&&M_{e}^{\mu\nu\alpha\beta\gamma}=\Gamma_{3}^{\mu\alpha\xi}(-k_{1},p_{+},k_{1}-p_{+})
D_{\xi\lambda}(k_{1}-p_{+})\Gamma_{4}^{\lambda\beta\nu\gamma}+[k_{1\leftrightarrow2};\mu\leftrightarrow\nu],\\
\nonumber
&&M_{f}^{\mu\nu\alpha\beta\gamma}=\Gamma_{3}^{\alpha\gamma\xi}(p_{+},k_{3},-k_{3}-p_{+})
D_{\xi\lambda}(k_{3}+p_{+})\Gamma_{4}^{\lambda\beta\nu\mu},\\
\nonumber
&&M_{g}^{\mu\nu\alpha\beta\gamma}=\Gamma_{3}^{\gamma\beta\xi}(k_{3},p_{-},-k_{3}-p_{-})
D_{\xi\lambda}(k_{3}+p_{-})\Gamma_{4}^{\lambda\alpha\nu\mu}.
\end{eqnarray}
Where $[k_{1\leftrightarrow2};\mu\leftrightarrow\nu]$ indicates
the
 crossing contributions of the initial photons.

 \indent With the above amplitudes, we can directly obtain the cross
section $\hat{\sigma}(\hat{s})$ for the sub-process
$\gamma\gamma\rightarrow W^{+}W^{-}B_{H}$ and the total cross
section at the $e^+e^-$ linear collider can be obtained by folding
$\hat{\sigma}(\hat{s})$ with the photon distribution function
which is given in Ref\cite{distribution}
\begin{eqnarray}
\sigma_{tot}(s)=\int^{x_{max}}_{x_{min}}dx_{1}\int^{x_{max}}_{x_{min}
x_{max}/x_1}dx_{2} F(x_{1})F(x_{2})\hat{\sigma}(\hat{s}),
\end{eqnarray}
where $s$ is the c.m. energy squared for $e^+e^-$ and the
subprocess occurs effectively at $\hat{s}=x_1x_2s$, and $x_i$ are
the fraction of the electron energies carried by the photons. The
explicit form of the photon distribution function $F(x)$ is
\begin{eqnarray}
\displaystyle F(x)=\frac{1}{D(\xi)}\left[1-x+\frac{1}{1-x}
-\frac{4x}{\xi(1-x)}+\frac{4x^2}{\xi^2(1-x)^2}\right],
\end{eqnarray}
with
\begin{eqnarray}
\displaystyle D(\xi)&=&\left(1-\frac{4}{\xi}-\frac{8}{\xi^2}\right)
\ln(1+\xi)+\frac{1}{2}+\frac{8}{\xi}-\frac{1}{2(1+\xi)^2},\\
\end{eqnarray}
with
\begin{eqnarray}
\xi=\frac{4E_0\omega_0}{m^2_e},
\end{eqnarray}
and $E_0$ and $\omega_0$ are the incident electron and laser light
energies. The energy $\omega$ of the scatered photon depends on
its angle $\theta$ with respect to the incident electron beam and
is given by
\begin{eqnarray}
\omega=\frac{E_0(\frac{\xi}{1+\xi})}{1+(\frac{\theta}{\theta_0})^2}.
\end{eqnarray}
Therefore, at $\theta=0$, $\omega=E_0\xi/(1+\xi)=\omega_{max}$ is
the maximum energy of the back-scattered photon and
$x_{max}=\frac{\omega_{max}}{E_0}=\frac{\xi}{1+\xi}$ is the
maximum fraction of energy carried away by the back-sacttered
photon.

 To avoid unwanted $e^+e^-$ pair production from
the collision between the incident and back-scattered photons, we
should not choose too large $\omega_0$. The threshold for $e^+e^-$
pair creation is $\omega_{max}\omega_0>m^2_e$, so we require
$\omega_{max}\omega_0\leq m^2_e$. Solving $\omega_{max}\omega_0=
m^2_e$, we find
\begin{eqnarray}
\xi=2(1+\sqrt{2})=4.8.
\end{eqnarray}
 For the choice $\xi=4.8$, we obtain $x_{max}=0.83$ and $D(\xi_{max})=1.8$.
 The minimum value for $x$ is determined by the production
threshold
\begin{eqnarray}
x_{min}=\frac{2M_W+M_{B_H}}{x_{max}s}.
\end{eqnarray}

Here we assume that both photon beams and electron beams are
unpolarized. We also assume that, on average, the number of the
back-scattered photons produced per electron is 1, i.e., the
conversion coefficient $k$ is equal 1.

\section{ Numerical results and conclusions}

 \hspace{0.6cm} In our calculations, we take $M_{W}=80.283$ GeV,
$v=246$ GeV,  $s^{2}_{W}=0.23$. The electromagnetic fine structure
constant $\alpha_e$ at certain energy scale is calculated from the
simple QED one-loop evolution formula with the boundary value
$\alpha=1/137.04$ \cite{Donoghue}. There are three free
parameters($f,c',\sqrt{s}$) involved in the production amplitudes.
Global fits to the electroweak precision data produce rather
severe constraints on the parameter spaces of the littlest Higgs
model. However, if we carefully adjust the $U(1)$ section of the
theory,
 the contributions to the electroweak observables can be reduced and
 the constraints become
 relaxed. The scale parameter $f=1-2$ TeV is allowed for the mixing parameters $c$ and
  $c'$ in the range of $0-0.5$ and $0.62-0.73$, respectively\cite{constraints}. The numerical results
  are summarized in Fig.2-4.

\indent  In general, the contributions of the littlest Higgs model
to the observables are dependent on the factor $1/f^{2}$. To see
the effect of the varying $f$ on the production cross section
$\sigma_{W^{+}W^{-}B_{H}}$, we plot $\sigma_{W^{+}W^{-}B_{H}}$ as
a function of $f(f=1-2$ TeV) for three values of the mixing
parameter $c'$ in Fig.2. Where we take $\sqrt{s}=800, 1500$ GeV as
the examples of the ILC c.m. energies. We can see that the cross
section fall sharply with $f$ increasing. The main reason is that
the production amplitudes vary depending on factor $1/f^2$.
Comparing the results of Fig.2.(a) and Fig.2.(b), we find that the
large $\sqrt{s}$ can enhance the cross section significantly
because there is no s-channel depression by the large $\sqrt{s}$.
For small value of $f$, the cross section can reach the level  of
a few $fb$ with $\sqrt{s}=800$ GeV and tens $fb$ with
$\sqrt{s}=1500$ GeV. The large $\sqrt{s}$ is favorable for the
detection of $B_{H}$.

\indent To study the effect of $c'$, in Fig.3, we show the cross
section as a function of $c'$ with $\sqrt{s}=1500$ GeV and $f=1$
TeV, 1.2 TeV, 1.5TeV, respectively.

 From Fig.3, one can see that the cross section decreases sharply with
 $c'$ increasing for $c'<\frac{\sqrt{2}}{2}$. However, for
$c'>\frac{\sqrt{2}}{2}$, the cross section increases with $c'$
increasing. For $c'=\frac{\sqrt{2}}{2}$, the value of the cross
section is zero. This is because the gauge boson self-couplings
$g_{W^{+}W^{-}B_{H}}$ and $g_{\gamma W^{+}W^{-}B_{H}}$ are
proportional to $s'c'(c'^{2}-s'^{2})$ and such couplings become
decoupled when $c'=\frac{\sqrt{2}}{2}$. On the other hand, for
small values of $f$ and $c'$, the cross section can reach a few fb
even tens fb. If we take integral luminosity
$\pounds=1000fb^{-1}$, there are about $10^3-10^4$ $W^+W^-B_H$
events to be produced. There will be a promising number of fully
reconstructible events to detect $B_{H}$. Furthermore, it is
possible to study the gauge boson self-couplings via
$\gamma\gamma\rightarrow W^{+}W^{-}B_{H}$ with large precision at
the ILC.

To illustrate the influence of the coupling $G$ defined in Eq.(22)
and $B_H$ mass on the cross section, in Fig.4, we show the plots
of $G^2$ and $M_{B_H}$ as a function of $c'$ with fixed $f=1$ TeV.
 The change of $B_H$ mass
$M_{B_H}$ can affects the phase space, but $M_{B_H}$ does not
change very much when $c'$ changes in the range $0.62-0.73$. So
the change of the cross section mainly depends on the coupling
$G$.

\indent  Since we are interested in final states where all the
gauge bosons are identified, the event rate is determined not only
by the total cross section, but also by the reconstruction
efficiency that depends on the particular decay channels of the
vector bosons. The efficiency for reconstruction of a $W^{\pm}$ is
over $60\%$\cite{efficiency}. To identify $B_H$ from the final
states, we also need to study its decay modes. The main decay
modes of $B_H$ are $e^+e^-+\mu^+\mu^-+\tau^+\tau^-,
d\bar{d}+s\bar{s}, u\bar{u}+c\bar{c}, Zh, W^+W^-$. The decay
branching ratios of these modes have been studied in reference
\cite{Han} which are strongly dependent on the $U(1)$ charge
assignments of the SM fermions. The most interesting decay modes
of $B_H$ should be $e^+e^-,\mu^+\mu^-$. This is because such
particles can be easily identified and the number of
$e^+e^-,\mu^+\mu^-$ background events with such a high invariant
mass is very small. So, a search for a peak in the invariant mass
distribution of the either $e^+e^-$ or $\mu^+\mu^-$ is sensitive
to the presence of $B_H$. In the SM, the same sufficient final
states can also be produced via $\gamma\gamma\rightarrow W^+W^-Z$
with $Z\rightarrow e^+e^-(\mu^+\mu^-)$, and the cross section is
over $10^2$ fb\cite{WWZ}. It should be very easy to distinguish
$B_H$ from $Z$ when we look at the invariant mass of the $e^+e^-$
or $\mu^+\mu^-$ pair because there might exist significantly
different $e^+e^-$($\mu^+\mu^-$) invariant mass distribution
between $B_H$ and $Z$. On the other hand, $B_H$ can also decay to
$W^+W^-, Zh$ and these bosonic decay modes are dominated by the
longitudinal components of the final-states. In general, the decay
branching ratios of $W^+W^-$ and $Zh$ are very small, but for the
favorable parameter spaces we might assume enough $W^+W^-$ and
$Zh$ signals to be produced with high luminosity. Such signals
would provide crucial evidence that an observed new gauge boson is
of the type predicted in the little Higgs models. For
$B_H\rightarrow Zh$, the final states are $l^+l^-b\bar{b}$. Two
b-jets reconstruct to the Higgs mass and a $l^+l^-$ pair
reconstructs to the Z mass. On the other hand, the decay mode $Zh$
involves the off-diagonal coupling $hZB_H$ and the experimental
precision measurement of such off-diagonal coupling is more easier
than that of diagonal coupling. So, the decay mode $Zh$ would
provide a better way to verify the crucial feature of quadratic
divergence cancellation in Higgs mass. The decay mode
$Z\rightarrow W^+W^-$ is of course kinematically forbidden in SM,
but the decay $h\rightarrow W^+W^-$ is the dominant decay mode of
the Higgs boson with mass above 135 GeV(one or both of the W
bosons is off-shell for Higgs mass below $2M_W$). We leave the
detailed study of such SM background to experimentalists.

The $B_H$ production mechanics has been studied at the
LHC\cite{LHC}, and such particle can also be produced by $e^+e^-$
collision. However, these production processes involve the fermion
couplings and these couplings suffer from large theoretical
uncertainty due to the arbitrariness of the fermion $U(1)$ charge
assignments. So, the reliable prediction of the $B_H$ signals via
these processes would be difficult. On the other hand, with the
small $e^+e^-B_H$ coupling and high energy s-channel suppression,
the s-channel $B_H$ production in $e^+e^-$ collisions would of
course be suppressed which makes $\gamma\gamma$ production mode
become even more important, specially for gaining an insight into
the gauge structure of the littlest Higgs model.

 \indent In summary, the new gauge bosons are the typical particles of the littlest Higgs model.
 With the mass in the scale of hundreds GeV, the $U(1)$ boson $B_H$ is the lightest one among these new
 gauge bosons. Therefore, such particle might provide a early signal of the littlest Higgs model
 at the ILC. Because the high c.m. energy can significantly enhance the cross
 sections of the triple gauge boson production processes, these processes become more important at the ILC.

 In this paper, we study a $B_H$ production process  associated with W boson pair
 via
$\gamma\gamma$ collision. It can be concluded that the cross
section is sensitive to the parameters $f, c',\sqrt{s}$ which make
the cross section vary from $10^{-1}$ to $10^1$ fb in most
parameter spaces allowed by the electroweak precision data. With
the favorable parameter spaces(the high c.m. energy, small $c'$
and $f$), the sufficient events can be produced to detect $B_H$.
On the other hand, if such gauge boson is observed at future
collider experiments, the precision measurement is need which
could offer the important insight for the gauge structure of the
littlest Higgs model and distinguish this model from alternative
theories.

\newpage
\begin{figure}[t]
\begin{center}
\epsfig{file=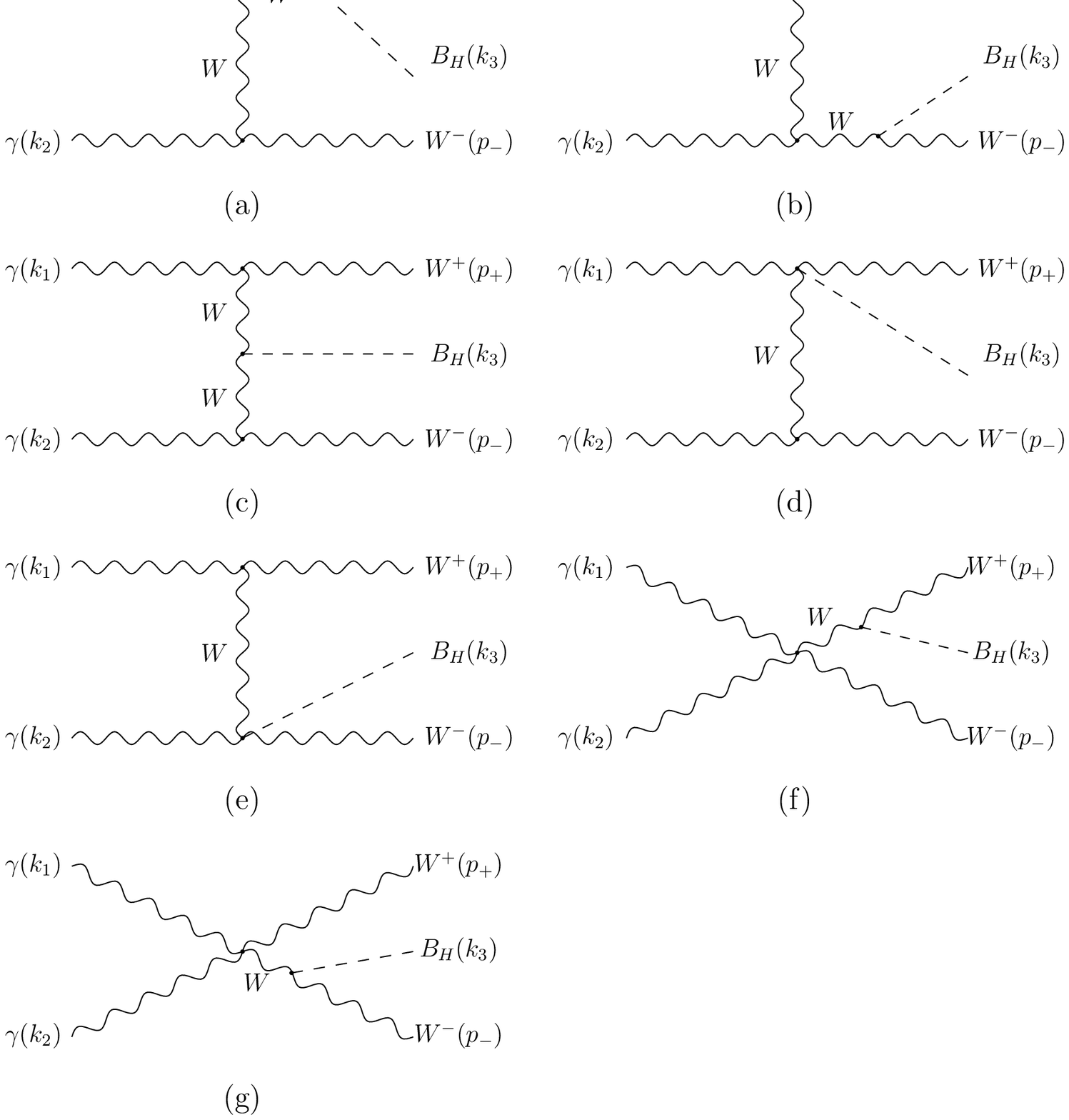,width=450pt,height=640pt} \vspace{-6.5cm}
\caption{\small The Feynman diagrams of the process
$\gamma\gamma\rightarrow W^{+}W^{-}B_{H}$ in the littlest Higgs
model.} \label{fig1}
\end{center}
\end{figure}
\newpage
\begin{figure}[ht]
\begin{tabular}{cc}
\scalebox{0.75}{\epsfig{file=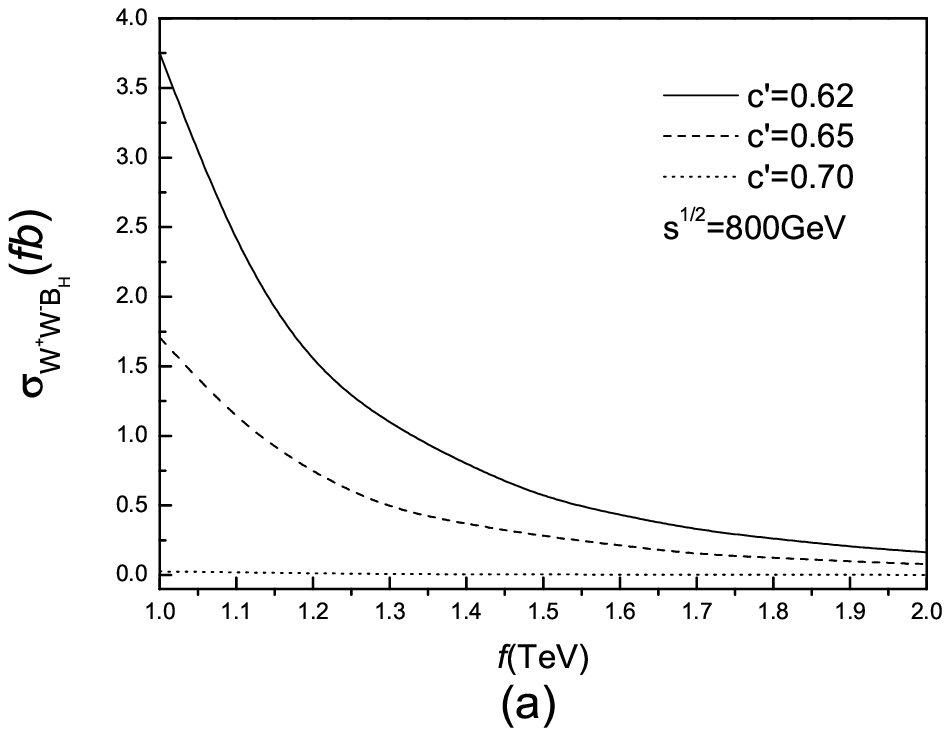}}\\
\end{tabular}
\begin{tabular}{cc}
\scalebox{0.75}{\epsfig{file=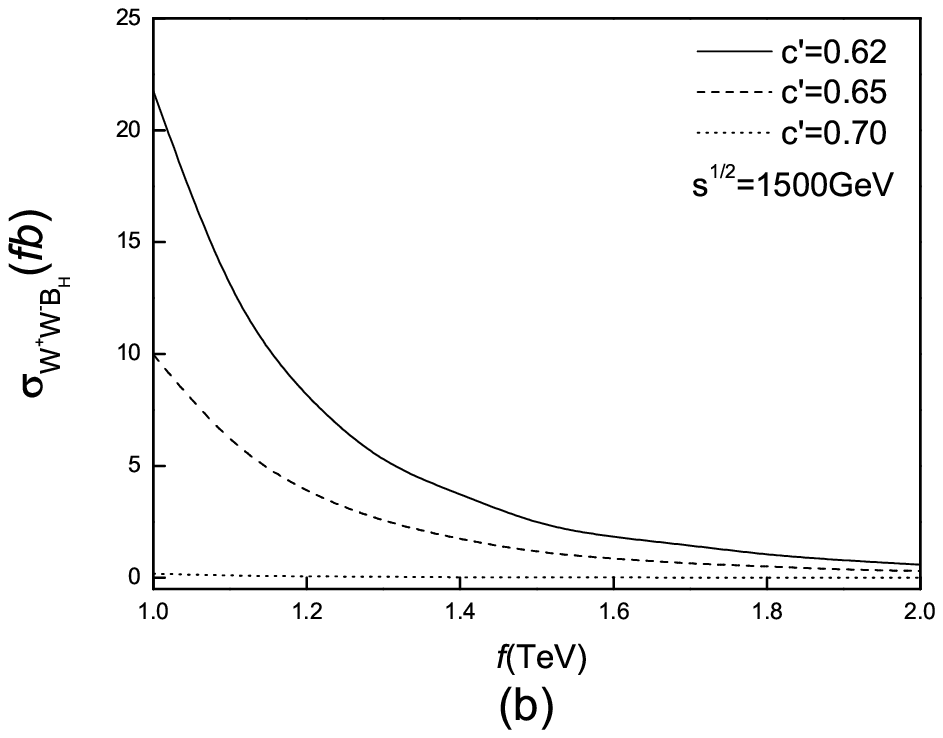}}\\
\end{tabular}
\caption{\small The production cross section of the process
$\gamma\gamma\rightarrow W^{+}W^{-}B_{H}$ as a function of the
scale parameter $f$ for three different values of the mixing
parameter $c'$, with $\sqrt{s}=800$ GeV, 1500 GeV, respectively.}
\end{figure}

\begin{figure}[hb]
\begin{tabular}{cc}
~~~~~~~~~~~~~~~~~~~~~~~\scalebox{0.85}{\epsfig{file=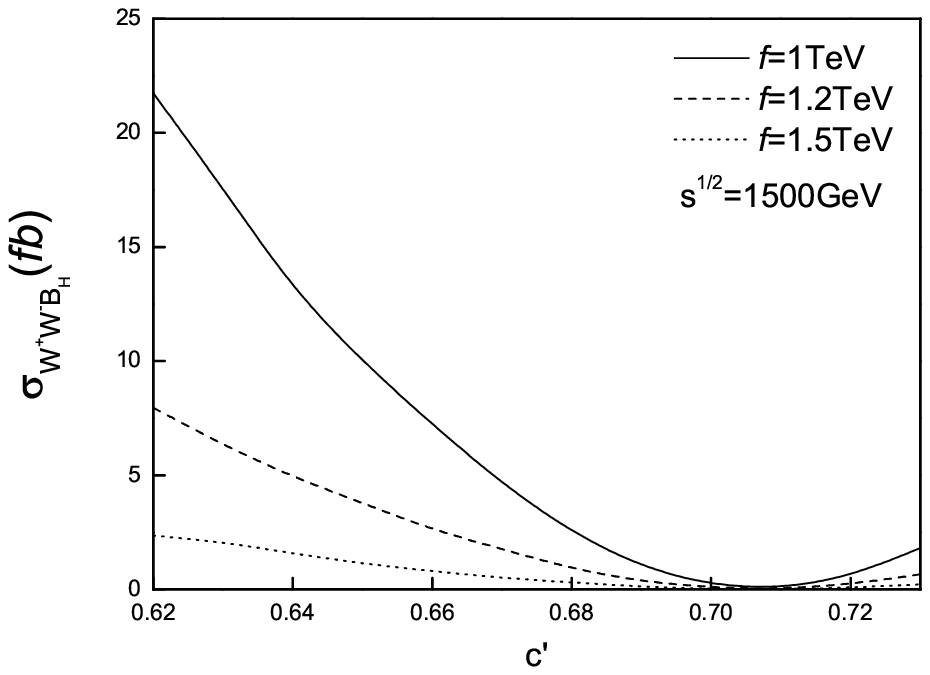}}
\end{tabular}
\caption{\small The production cross section of the process
$\gamma\gamma\rightarrow W^{+}W^{-}B_{H}$ as a function of the
mixing parameter $c'$  for three different values of the scale
parameter $f$ with $\sqrt{s}=1500$ GeV.}
\end{figure}

\newpage
\begin{figure}[h]
\begin{tabular}{cc}
\scalebox{0.75}{\epsfig{file=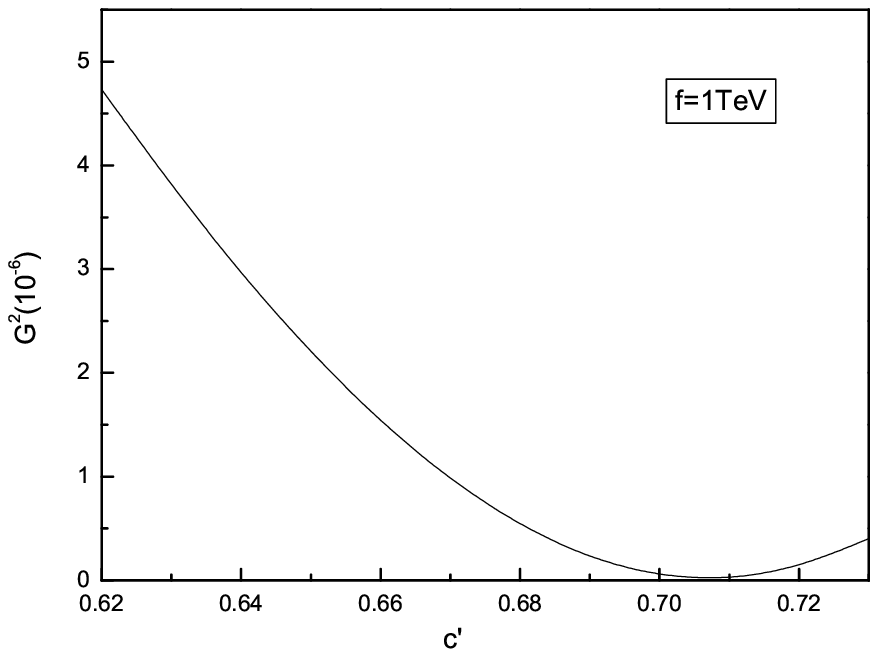}}\\
\end{tabular}
\begin{tabular}{cc}
\scalebox{0.75}{\epsfig{file=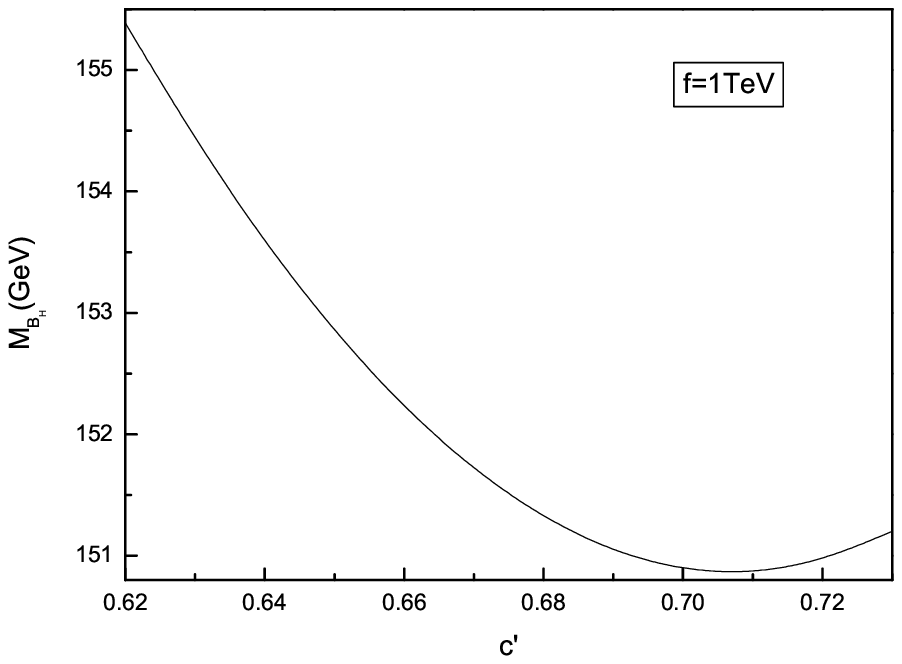}}\\
\end{tabular}
\caption{\small The plots of $G^2$ and $M_{B_H}$ as a function of
$c'$ with fixed $f=1$ TeV.}
\end{figure}

\end{document}